

\magnification 1200

\overfullrule 0pt

\parskip 0pt plus 1pt
\parindent=35pt
\advance\vsize by -0,9 true cm\advance\voffset by 0,9 true cm
\advance\hsize by -1,36 true cm\advance\hoffset by 1,26 true cm

\font\eightrm=cmr8
\font\gross=cmcsc10

\hfill

\topinsert\vskip 1 true cm\endinsert
\centerline{\bf ``Elements of Reality" and the Failure of the Product
Rule }

\vskip 1.5truecm
\centerline{{\gross Lev Vaidman}}

\centerline{{\eightrm School of Physics and Astronomy }}
\centerline{{\eightrm
Raymond and Beverly Sackler Faculty of Exact Sciences }}
\centerline{{\eightrm Tel-Aviv University, Tel-Aviv, 69978 ISRAEL}}

\vskip 1true cm
\centerline{\gross Abstract}
\vskip 10pt
{\advance\baselineskip by -2pt
\par{\narrower\noindent
{\eightrm
The concept of ``elements of reality" is analyzed within the framework
of quantum theory. It is shown that elements of reality
fail to fulfill the product rule. This is the core of recent
proofs of the impossibility of a Lorentz-invariant interpretation of
quantum mechanics. A generalization and extension of the concept of
elements of reality is presented. Lorentz-invariance is
 restored by giving up the product rule. The consequences of giving up
the ``and" rule, which must be abandoned together with the product
rule,
are discussed. }\par}}

\vskip 1truecm

\noindent
{\bf 1. ``Elements of Reality"}
\vskip .1cm

One of the main tasks of physics is to give a
(mathematical) description of reality.  At the beginning of the century
physicists thought that they were close to completing this task.
Newtonian mechanics and classical electrodynamics explained very well
most of the observed phenomena.
The theory of relativity showed that ``reality" is much more
bizarre than the Laplacian mechanics in Cartesian space, but
physics is still capable of describing it.  Experiments showed,
however, relativistic
classical physics does not describe reality.
Experiments are in extremely
good agreement with another theory: quantum
mechanics.

 If quantum mechanics is the correct physics theory,
 then it is very difficult to see how physics can
describe reality.  Einstein, Podolsky, and Rosen$^1$ (EPR)
 argued that quantum mechanics cannot give a complete
description of reality and Bell$^2$ even showed that it is impossible
to have a (local) complete description of reality which is compatible
with quantum mechanics.  Essentially, he showed that using our common
``classical" concepts we cannot describe the world as it is.

One way out (close to Bohr's position) is to postulate that reality is
what we find in our measurements, and that quantum
mechanics
is a mathematical tool for calculating probabilities for these
results. However, if you  believe that the
moon is there even if nobody looks at it,
and if you have the same  belief about an electron as you have
about the moon,
you must search for a different
definition of physical reality.

Beyond the limitations on local realistic theories demonstrated by
Bell, recently, several authors$^{[3-6]}$ claimed that there are even
stronger restrictions on realistic theories: it is also impossible to
build a realistic Lorentz-invariant theory consistent with quantum
mechanics.  Technically their results are correct, but we disagree
with their conclusions.  A crucial issue is what we understand as an
``element of reality".  Since classical physics is incorrect, it is
not surprising that common classical concepts are not appropriate for
a description of physical reality.  As the theory of relativity taught
us to change radically our concepts of space and time, quantum
mechanics leads us to alter our concept of reality.  We will show that
the claim that Lorentz-invariance conflicts with realism relies on an
assumed classical property of elements of reality, which does not hold
in our (quantum) world\rlap.$^7$

\vskip .4cm
\noindent
{\bf 2. The Product Rule}

 The proofs of the impossibility of realistic Lorentz-invariant
quantum theory used the
{\it product rule}:
 ~If $A$ and $B$ commute, $A=a$ and $B=b$, then $AB
= ab$.

In fact, Fine and Teller\rlap,$^8$\ based on Bell's paper\rlap,$^2$
showed already in 1977 that one cannot construct a hidden variable
(i.e.
realistic) theory, compatible with quantum mechanics, which obeys
the product rule.   It was shown even more explicitly  in the works of
Peres\rlap.$^{9-10}$\ However, the product rule holds
 in standard quantum
mechanics, and the recent definitions of elements of reality go only
half
way between quantum mechanics and local hidden variables, so
it is not obvious what is the status of the product rule in this
case.  We claim that according to the recent definition,
the product
rule {\it should not} be used
in
the situations which have been considered.
 We  present simple examples
in which the product rule clearly fails.  But, let us
first review the product rule in usual
situations of quantum mechanics, where it certainly holds.

In every textbook of quantum mechanics we can find a condition
for simultaneous measurability of variables $A$ and $B$; the
corresponding operators must commute:
$$[A,B] = 0. \eqno(1)$$
Commutativity of the operators $A$ and $B$ is a strong sufficient
condition; for a given quantum state $|\Psi \rangle$, it is
sufficient to have commutativity with respect to that state:
$$[A,B] |\psi \rangle = 0. \eqno(2)$$
The commutativity condition (2) is
a necessary and
sufficient
condition for
simultaneous measurability of $A$ and $B$.  If the operators $A$ and
$B$ do not commute, the measurement of one disturbs the outcome of the
other. For example, consider a
standard
measuring procedure$^{11}$
with an interaction Hamiltonian given
by
$$H = g(t) p A.\eqno(3)$$
Here $p$ is a canonical momentum of the measuring device; the
conjugate position $q$ corresponds to the position of a pointer on the
device.  The coupling $g(t)$ is non-zero for a short
time  interval, and during the
measurement we obtain (in the Heisenberg picture)
$${dB \over dt}=i [H,B] =ig(t) p [A,B].
\eqno(4)$$
 Thus, commuting operators
are measurable without mutual disturbance, while
non-commuting operators disturb one another.

 If $A$ and $B$ commute, and if  we know that at a given moment
a  measurement of $A$ (if performed) must yield $A=a$ while a
measurement
of $B$ (if performed) must yield $B=b$, we can safely claim that the
product a
 $AB$ is also known and equal $ab$.
We repeat this well-known fact because, surprisingly, it is not
true when we consider a pre- and post-selected quantum system.

\vskip .4cm
\noindent
{\bf 3. The Pre- and Post-Selected Ensemble}
\vskip .1cm

To define a
pre- and post-selected quantum system, we
consider a quantum system at time $t$.  For simplicity we let
the free Hamiltonian be zero.  At time $t_1 <t$ the system is
prepared
in a quantum state $|\Psi_1\rangle$, and at a time $t_2 >t$ a
measurement is performed and the system is found in the
state $|\Psi_2\rangle$.  We ask about possible measurements
at time $t$. Suppose
 $A$ is measured at time $t$.  If either $|\Psi_1\rangle$
or $|\Psi_2\rangle$ is an eigenstate of $A$, then clearly the outcome
of the measurement is determined; it is the corresponding
eigenvalue
of $A$. Measuring the commuting operator  $B$ before, after,
or
even during the measurement of $A$ does not, in principle, disturb the
measurement of $A$.  However, for a pre- and post-selected quantum
system it might be that the result of measuring $A$ is
certain, even if neither $|\Psi_1\rangle$ nor $|\Psi_2\rangle$ is an
eigenstate of $A$.  In this case  a measurement at any
time between  $t_1$ and $t_2$ of certain operators
commuting with $A$ {\it invariably disturbs the $A$-measurement.}

A simple example is the setup proposed by Bohm for analyzing the EPR
argument: two separate spin-1/2 particles prepared, at time $t_1$, in
a singlet state $$|\Psi_1 \rangle = {1\over {\sqrt 2}} (|\uparrow_1
\downarrow_2\rangle - |\downarrow_1 \uparrow_2 \rangle ). \eqno(5)$$
At time $t_2$ measurements of ${\sigma_1}_x$ and ${\sigma_2}_y$ are
performed and certain results are obtained.  If at time $t$, $ t_1 < t
< t_2$, a measurement of ${\sigma_1}_y$ is performed (and if this is
the only measurement performed between $t_1$ and $t_2$), then the
outcome of the measurement is known with certainty: ${\sigma_1}_y(t) =
-{\sigma_2}_y(t_2)$.  If, instead, only a measurement of
${\sigma_2}_x$ is performed at time $t$, the result of the measurement
is also certain: ${\sigma_2}_x(t) =- {\sigma_1}_x(t_2)$.  The
operators ${\sigma_1}_y$ and ${\sigma_2}_x$ obviously commute, but
nevertheless, measuring ${\sigma_2}_x(t)$ clearly disturbs the outcome
of the measurement of ${\sigma_1}_y(t)$: it is not certain anymore.

Measuring the product  ${\sigma_1}_y {\sigma_2}_x$,  is, in
principle, different from the measurement of both ${\sigma_1}_y$ and
${\sigma_2}_x$ separately.
In our example the outcome of the measurement of the product {\it
is} certain, but it does not equal the product of the results
which must come out of the measurements of
${\sigma_1}_y$ and ${\sigma_2}_x$ when every one of them is performed
without the other. To measure the product ${\sigma_1}_y
{\sigma_2}_x$
we may write it
as a modular sum, ${\sigma_1}_y {\sigma_2}_x = (
{\sigma_1}_y + {\sigma_2}_x){\rm mod4} - 1$. It has been shown$^{12}$
 that nonlocal operators such as $({\sigma_1}_y +
{\sigma_2}_x){\rm mod4}$ can be measured using
 solely local interactions.

In order to find out the results of the measurements
we can use the generalization of the formula of
Aharonov, Bergmann, and Lebowitz$^{13}$
(ABL) for
calculating probabilities for the results of an intermediate
measurement performed on a pre- and post-selected system.
  If the initial
state is $|\Psi_1\rangle$ and the post-selected
state is
$|\Psi_2\rangle$, then the probability for an intermediate measurement
of $A$ to yield  $A=a_n$ is given
by$^{14}$
$$
{\rm prob}[A=a_n ] = {{| \langle \Psi_2| {\bf P}_{A=a_n}
|\Psi_1\rangle |^2} \over {\sum_k
| \langle \Psi_2| {\bf P}_{A=a_k}
|\Psi_1\rangle
 |^2}}~~~~.
\eqno(6)$$
where the sum is over all eigenvalues of $A$ and ${\bf P}_{A=a_k}$ is
the projection operator onto the subspace with eigenvalue $a_k$.
The formula immediately yields probability 1 when $|\Psi_1\rangle$ or
$|\Psi_2\rangle$ is an eigenstate, but it also can yield 1 when
neither of the states is an eigenstate, as we now show.

The state $|\Psi_1 \rangle$ is given by Eq.~(5).  Suppose the results
of the post-selection measurements are ${\sigma_1}_x = 1$ and
${\sigma_2}_y = 1$.  Then the state $|\Psi_2 \rangle = |{\uparrow_1}_x
{\uparrow_2}_y \rangle $. To predict the outcome of a measurement of
${\sigma_1}_y$ we have to use the projection operators $ {\bf
P}_{[{\sigma_1}_y = 1]} = |{\uparrow_1}_y \rangle \langle
{\uparrow_1}_y |$ and $ {\bf P}_{[{\sigma_1}_y = -1]} =
|{\downarrow_1}_y \rangle \langle {\downarrow_1}_y |$.  Applying
formula (6), we indeed obtain prob[${\sigma_1}_y = -1] = 1$.  In the
same way we obtain prob[${\sigma_2}_x= -1] = 1$.  For calculation of
the probabilities of the measurement of the product ${\sigma_1}_y
{\sigma_2}_x$ we use the projection operators $$\eqalign{ {\bf
P}_{[{\sigma_1}_y {\sigma_2}_x = 1]} =& |{\uparrow_1}_y {\uparrow_2}_x
\rangle \langle {\uparrow_1}_y {\uparrow_2}_x| + |{\downarrow_1}_y
{\downarrow_2}_x \rangle \langle {\downarrow_1}_y
{\downarrow_2}_x|,\cr {\bf P}_{[{\sigma_1}_y {\sigma_2}_x = -1]} =&
|{\uparrow_1}_y {\downarrow_2}_x \rangle \langle {\uparrow_1}_y
{\downarrow_2}_x| + |{\downarrow_1}_y {\uparrow_2}_x \rangle \langle
{\downarrow_1}_y {\uparrow_2}_x|. \cr}\eqno(7)$$ Then Eq.~(6) yields
prob[$ {\sigma_1}_y {\sigma_2}_x= 1] = 0$, contrary to the product
rule, which requires $ {\sigma_1}_y {\sigma_2}_x= 1$ with probability
1. It follows that the value of the product $ {\sigma_1}_y
{\sigma_2}_x$ is certain, but it equals $-1$.

Another striking example was discussed by Albert, Aharonov and
D'Amato\rlap.$^{15}$\ Consider a particle which can be located in
one  of
three boxes. We denote the state of the particle when it is in box
$i$ by $|i\rangle $. At time $t_1$ the particle is prepared in the
state
$$|\Psi_1\rangle = {1 \over {\sqrt{3}}} (|1\rangle + |2\rangle +
|3\rangle ).\eqno(8)$$
At time $t_2$ the particle is found to be in the state
$$|\Psi _2\rangle = {1\over {\sqrt{3}}}(|1\rangle + |2\rangle -
|3\rangle).\eqno(9)$$
We assume that in the time interval $[t_1, t_2]$ the Hamiltonian is
zero.  Then, if at time $t$ between $t_1$ and $t_2$ we open box
1, we are certain to find the particle in box 1; and if we
open box 2 instead, we are certain to find the particle in box 2.
Nevertheless, if we open both of them, we might not see the particle
at all.  We can obtain these results by straightforward
application of the formula (6).  Opening box $i$ corresponds to
measuring the projection operator $|i\rangle \langle i|$.  Opening two
boxes is equivalent to opening the third box, and, therefore,
corresponds
to measuring the projection operator on the state in the third box.

We have shown that for a pre- and post-selected quantum system
it might happen that the operators corresponding to two observables
$A$ and $B$ commute, $[A,B] = 0$, but measuring $B$ invariably
disturbs the results of the measurement of $A$.  Therefore, for a pre-
and post-selected quantum system one cannot apply a ``product rule"
that asserts that if measurements of $A$ and $B$ yield
 $A=a$ and $B=b$ with certainty, then a measurement of $AB$ yields
$ab$.  In fact, the value of $AB$ might also be certain, but not equal
to $ab$.

\vskip .4cm
\noindent
{\bf 4. Realistic Lorentz-invariant
Interpretation of Quantum Mechanics}
\vskip .1cm

We now turn to  the arguments against the possibility of a
realistic Lorentz-invariant interpretation of quantum
mechanics\rlap.$^{3-6}$\ The starting point of these
arguments was the definition of elements of reality and the principle
of Lorentz invariance.  In contrast to the usual EPR-type argument, no
locality principle, forbidding an action at a distance, was assumed.
The adopted  definitions are:
 \item{(i)} {\it Element of
reality} (Redhead$^{16}$):``If we can predict with
certainty, or
at any rate with probability one, the result of measuring a physical
quantity at time $t$, then at the time $t$, there exists an element of
reality corresponding to this physical quantity and having a value
equal to the predicted measurement result."

\item{(ii)} {\it The Principle of Lorentz invariance}: ``If an element
of reality corresponding to some Lorentz-invariant physical quantity
exists and has a value within space-time region R with respect to one
space-like hypersurface containing R, then it exists and has the
same value in R with respect to any other hypersurface containing
R."

\vskip .1cm
In the usual EPR argument an element of reality corresponding to an
outcome of a measurement is fixed by the mere {\it possibility} of
inferring the outcome from measurements in a causally disconnected
region.  In contrast, since the present approach does not assume
locality, elements of reality are fixed only by {\it actual}
measurements.

\break

\vskip .4cm
\noindent
{\bf 5. The Proof of Clifton, Pagonis, and  Pitowsky}
\vskip .1cm

The argument due to Clifton$^4$ is based on the modified
Greenberger-Horne-Zeilinger$^{17}$ (GHZ) setup for demonstrating the
nonexistence of local hidden variables.  Three spin-1/2 particles,
located in the corners of a very large triangle, move fast in
directions pointing out of the center of the triangle.  At time $t_1$
(in the rest frame) the particles are prepared in the state $$ |\Psi_1
\rangle = |GHZ \rangle = {1\over {\sqrt{3}}} ({|\uparrow_1}_z
{\uparrow_2}_z {\uparrow_3}_z \rangle - {\downarrow_1}_z
{\downarrow_2}_z {\downarrow_3}_z \rangle). \eqno(10)$$ At time $t_2$
the spin components in $x$-direction are measured on all particles and
the results ${\sigma_i}_x = x_i$ are obtained.  Consider now some
possible measurements performed on the particles at a time $t$, $t_1 <
t < t_2$.  For each of the three observers who perform the
${\sigma_i}_x$ measurements, the measurements on the other particles
(at time $t$ in the rest frame) are performed {\it after} his
${\sigma_i}_x$ measurement, and he can predict (each in his Lorentz
frame) the following result with certainty: $$ {\sigma_2}_y
{\sigma_3}_y = x_1 \eqno(11a)$$ $$ {\sigma_1}_y {\sigma_3}_y = x_2
\eqno(11b)$$ $$ {\sigma_1}_y {\sigma_2}_y = x_3 \eqno(11c)$$ Eqs.
(11a-c) represent elements of reality in space-time regions
corresponding to the respective Lorentz frames.  The principle of
Lorentz invariance yields that these are also the elements of reality
in the rest frame.  Multiplying Eqs.  (11b) and (11c) we obtain:
$${\sigma_1}_y^2 {\sigma_3}_y {\sigma_2}_y = x_2 x_3 \eqno(12)$$
Taking in account that ${\sigma_1}_y^2 = 1$, we conclude that $x_1 =
x_2 x_3 $. This conclusion, however, contradicts quantum mechanics: in
the GHZ state $x_1 x_2 x_3 = -1$.

Pitowsky and Clifton {\it et al.} obtain their elements of reality as
{\it
predictions} of different observers, but their argument holds only
when they consider the predictions of all observers.  However, there
is no Lorentz observer for which all the predictions are inferences
from the past toward the future: at least some of the inferences must
be {\it retrodictions}.  In fact, we have a quantum system on which
two complete measurement are performed in succession, and claims about
elements of reality apply to times between these two measurements.
Here, the GHZ state is prepared initially and measurement of the
$x$-components of spin for all particles determines the final state.
The discussion at the beginning of this Letter thus applies.

The state $|\Psi_1 \rangle$ is given by Eq.~(8), while $|\Psi_2
\rangle = | x_1, x_2, x_3 \rangle$, i.e., the state with certain
$x$-components of spin.  The operators to be considered between these
two states are ${\sigma_1}_y {\sigma_3}_y,~ {\sigma_1}_y
{\sigma_2}_y,$ and ${\sigma_2}_y {\sigma_3}_y $. The formalism,
Eq.~(6), yields (as it should) the probability 1 for the outcomes
given by Eqs.  (11a-c).  But it also shows that the measurements of
commuting operators $ {\sigma_1}_y {\sigma_3}_y,$ and ${\sigma_1}_y
{\sigma_2}_y $ disturb each other.  Eq.~(6) yields that the
probability to find both results (11b) and (11c), {\it when measured
together} is just $1\over4$.  Again, measuring the product differs
from measuring both of the operators separately, and the probability
of finding $({\sigma_1}_y {\sigma_3}_y)( {\sigma_1}_y {\sigma_2}_y) =
x_2 x_3$ is zero since the outcome is given by Eq.~(11a).

\vskip .4cm
\noindent
{\bf 6. The Proof of Hardy}
\vskip .1cm

The example of Hardy$^5$ involves just two particles, an
electron and a positron in two entangled setups of the type proposed
by Elitzur and Vaidman$^{18}$
(EV) for interaction-free
measurements.
 Each EV setup is a Mach-Zehnder interferometer
tuned to yield zero counts at a detector $D_1$ unless a point $\cal P$
belonging to one arm of the interferometer is not free.  The ``click"
of the detector $D_1$, after sending just one particle, yields that
the point $\cal P$ is not empty, without disturbing the object at
$\cal P$.  In Hardy's example the point $\cal P$ is common to the two
EV setups.  One EV device tests the point $\cal P$ with a single
electron, while the other tests the same point $\cal P$ with a single
positron.  If both electron and the positron come to the point $\cal
P$ together then they annihilate, and it might happen that both
devices yield that the point $\cal P$ is not empty, i.e., detectors
$D_1$ of both the electron and the positron interferometers ``click".
Let us assume this outcome.  Now, consider a Lorentz frame in which
the observer of the electron EV device is the first to obtain a
``click".  She infers
 that the {\it positron} was at $\cal P$.
In fact, she {\it retrodicts}, since the events
she infers were in her absolute past.  (Hardy is able to discuss the
observer's {\it predictions} by considering the question: ``Is the
particle in the arm of the interferometer which includes $\cal P?$"
instead of the question: ``Is the particle at $\cal P$?"  See also
Ref. (6).)  In another
Lorentz frame, however, the observer of the positron EV device is the
first to obtain the result.  He deduces that the {\it electron} was at
$\cal P$.  The principle of Lorentz invariance yields that there are
two elements of reality: the electron at $\cal P$ and the positron at
$\cal P$.  The product rule here is very natural: if the electron is
at $\cal P$ and the positron is at $\cal P$ then the electron and the
positron are at $\cal P$.  The latter, however, leads to
contradiction: the particles at $\cal P$ must annihilate and cannot be
detected by either observer.

Hardy's example also involves pre- and post-selection.  Here,
the pre-selection is the preparation of the electron-positron state,
while the post-selection is the detection of electron and positron at
detectors $D_1$.  Thus, we can apply the ABL formalism.  However,
in this case the free Hamiltonian is not zero; it describes the
interaction of
the electron and the positron with beam splitters and mirrors as well
as their annihilation at $\cal P$.
Therefore, the state $|\Psi_1 \rangle$ in the formula (6) must be the
initial state evolved forward in time until $t$, the time when one of
the particles reaches the point $\cal P$; while the state $|\Psi_2
\rangle$ must be obtained by evolving the final state backward in time
until $t$.  Straightforward calculation shows that Eq.~(6) reproduces
Hardy's result: if one observer tests: ``Was the electron at $\cal
P$?" her result must be ``yes"; if the other observer looks for the
positron at $\cal P$, his answer must be ``yes" too (but if both of
them makes these measurements, each observer will obtain ``yes" with
probability $1\over3$, and they will never obtain ``yes" together).
Here too, the operator considered by Hardy is the product of two
projection operators, and its measurement is not equivalent to two
simultaneous measurements, one testing for electron at $\cal P$ and
another testing for positron at $\cal P$.  The measurement of the
product can be implemented by observing photons due to
electron-positron annihilation.  Formula~(6) yields
probability zero to obtain the product equal 1, in contrast to
probability one obtained from the product rule.

\vskip .4cm
\noindent
{\bf 7. Inconsistency in Applying the Product Rule}
\vskip .1cm

We believe that Redhead's definition of elements of reality is a
plausible one.  It does not lead to contradiction with Lorentz
invariance if we do not adopt the product rule.  But in the light of
the discussion above, it is clear that the product rule is
incompatible
with Redhead's definition.  The elements of reality are inferred on
the assumption that there are no measurements disturbing their values.
Clifton Pagonis and  Pitovsky$^4$ state explicitly: ``For our
argument, we shall assume that no such intervening measurements take
place\rlap."\  But as we showed, measurements of the operators they
consider
{\it do} interfere with each other.  So, it is inconsistent with the
definition of the elements of reality to apply the product rule.
(Note that the product rule and its generalization to any
function
of commuting operators are widely used in no-hidden-variables
theorems\rlap.$^{19}$\
 It is
valid in all cases when no retrodiction is involved.) If it is an
element of reality that $A=a$ and it is an element of reality that
$B=b$, it does not follow that $AB =ab$ is an element of reality.  It
might be that the product $AB$ has a certain value and, therefore, is
an element of reality in the Redhead's sense, but it need not equal
$ab$.

In fact, this happens in all the examples we considered.  In the first
example we have elements of reality ${\sigma_1}_y = -1$, ${\sigma_2}_x
= -1$, and the product is also an element of reality, but
${\sigma_1}_y {\sigma_2}_x = -1$.  In the second example {\bf P}$_{1}
= 1$, {\bf P}$_{2} = 1$, but {\bf P}$_{1}${\bf P}$_{2} = 0$, where
{\bf P}$_{1}$, {\bf P}$_{2}$ are projection operators on the states
``the particle in the box 1" and ``the particle in the box 2"
respectively.  In the Pitowsky example the elements of reality are $
{\sigma_1}_y {\sigma_3}_y = x_2$, $ {\sigma_1}_y {\sigma_2}_y = x_3$,
and the product, $ ({\sigma_1}_y {\sigma_3}_y)({\sigma_1}_y
{\sigma_2}_y) = {\sigma_2}_y {\sigma_3}_y = x_1 $, but nevertheless
$x_2 x_3 \neq x_1$, ($x_2 x_3 = - x_1$).  In Hardy's example {\bf
P}$_{e^-} = 1$, {\bf P}$_{e^+} = 1$, but {\bf P}$_{e^-}${\bf P}$_{e^+}
= 0$, where {\bf P}$_{e^-}$, {\bf P}$_{e^+}$ are projection operators
on the states ``an electron at $\cal P$" and ``a positron at $\cal P$"
respectively.

Clifton Pagonis and  Pitovsky felt that the conclusions about the
impossibility
of constructing a realistic Lorentz-invariant quantum theory are too
strong.  They proposed a variety of ways to circumvent
these arguments, in particular, by {\it rejecting} elements of reality
corresponding to ``incompatible measurement context". It is possible
to deal with the failure of the product rule along these lines, but
we believe that the most natural way is to give up the product
rule. All the examples we considered show that
 the product rule fails to be true.

\vskip .4cm
\noindent
{\bf 8. Elements of Reality of the Pre- and Post-Selected Quantum
System}
\vskip .1cm

Giving up the product rule allows us to extend the concept of elements
of reality.  Since we anyway consider circumstances in which
retrodictions are involved, we may include retrodictions
fully and give them
the same status as to predictions.  In the examples presented
here, predictions were applied to future events as well as to
space-like separated events, while
retrodictions were applied only to space-like separated events.  We
propose to apply  retrodiction to the past also.
The Redhead definition of elements of reality continues to hold, with a
minor change
of ``predict" to ``infer\rlap."\  Then, in the case of two
spin-1/2 particles, the observer who measures ${\sigma_1}_x
(t_2) = 1 $ not only infers that ${\sigma_2}_x(t) = -1$ but also that
${\sigma_1}_x(t) = 1$. So we can add to the list of elements of
reality at time $t$ also ${\sigma_1}_x = 1$ and ${\sigma_2}_y = 1$.

According to the definition, the element of reality exists {\it
whether or not} the inference is actually verified.
Recently were introduced {\it weak measurements}$^{20}$
 which might
support this definition.  Weak measurements test elements of reality
almost without disturbing the quantum system.  They refer to ensembles
of pre- and post-selected systems.  Each system in the ensemble is
practically undisturbed by the interaction with the measuring device
(which is a standard but very weakly coupled measuring device), but
measurement of each system yields almost no information.  However,
collecting results across the ensemble, we find a result called {\it
weak value}. If the system was pre-selected in a state
$|\Psi_1\rangle$ and was post-selected in a state $|\Psi_2\rangle$,
then any weak enough measurement of any variable A yields its weak
value

$$ A_w \equiv {{\langle \Psi_2 | A | \Psi_1 \rangle}
\over {\langle \Psi_2 |\Psi_1 \rangle}} .\eqno(12)$$

It has been shown$^{14}$ that whenever there exists an element of
reality, its value is the weak value.  For dichotomic variables an
``inverse" theorem$^{14}$ is also true: {\it if the weak value
is equal to an eigenvalue, then it is an element of reality} (i.e., a
measurement has to yield this value).  In all our examples we consider
dichotomic variables, so we can obtain our results via the simpler
calculation of the weak value (12) rather than via
Eq.~(6).

We can define {\it weak elements of reality}.  Weak values of physical
variables (i.e. the outcomes of weak measurements) are weak
elements of reality.  The elements of reality of Redhead (with
``infer" instead of ``predict") are subset of weak elements of
reality.  In contrast to the reality of Redhead,
``weak reality"
is defined in all situations.  There are numerous situations in
which
a quantum system has no elements of reality at all in the sense of
Redhead  (namely, when mixed states are involved).

Another attractive property of weak elements of reality is
the sum rule:
if $A = B + C$ then $A_w = B_w + C_w$.  The sum rule is valid even for
noncommuting variables. Despite  this parallel with classical physics,
weak elements of reality  might be very unusual.  For
example, for the particle in the three boxes there are the following
weak
elements of reality: there is 1 particle in box 1, there is 1 particle
in box 2, there is $-1$ particle in box 3!  If we
weakly measure the number of particles in the boxes (using a  pre-
and
post-selected ensemble of triplets of boxes), say, by
measuring
the pressure on the walls of the boxes, then we will find pressure
corresponding to a particle in each of the first two boxes and the
negative of the same value in the third box\rlap.$^{14}$\

Although the sum rule holds for weak elements of reality,
 the product
rule fails even for commuting variables: it is easy to see from the
definition (12) that  $A =
B C$ does not imply $A_w = B_w C_w$. It has to be so because
there is
failure of the product rule at least for the subset of weak elements
of reality, Redhead's elements of reality. This is a somewhat
surprising result
for the example  we have considered of the two spin-1/2 particles.
Even weak (supposedly undisturbing) measurement of the product
${\sigma_1}_y {\sigma_2}_x$ will be different from the product of the
outcomes of the weak measurements of
${\sigma_1}_y$ and ${\sigma_2}_x$. Indeed, $({\sigma_1}_y)_w = -1$,
$({\sigma_2}_x)_w = -1$, and $({\sigma_1}_y {\sigma_2}_x)_w = -1$,
therefore, $({\sigma_1}_y)_w ({\sigma_2}_x)_w \neq
({\sigma_1}_y {\sigma_2}_x)_w$.

\vskip .4cm
\noindent
{\bf 9. The Failure of the ``And" Rule}
\vskip .1cm

Closely connected to the failure of the product rule is the failure of
the {\it ``and" rule}: if $A=a$ is an element of reality and if $B=b$
is
an element of reality, it does not follow that $\{A=a$ and $B=b\}$ is
an element of reality.  Formally, one can consider the projection
operator on a space of states characterized by $A=a$ and the
projection operator on a space of states characterized by $B=b$; then
the product of these projection operators corresponds to the space of
states characterized by $\{A=a$ and $B=b\}$.  The failure of the
product rule for this case implies the failure of the ``and"
rule.

In fact, two of the examples presented are much more transparent when
we consider the ``and" rule instead of the product rule.  In the case
of a particle in three boxes we have: $\{$the particle is in box 1$\}$
 is an
element of reality,
$\{$the particle is in box 2$\}$ is an element of reality,
but $\{$the particle is in box 1 and the particle is in box 2$\}$
 is not
an element of reality.  In Hardy's example
$\{$the electron at $\cal P$ $\}$
is an element of reality,
$\{$the positron is at $\cal P$ $\}$ is an element
of reality, but
$\{$the electron and the positron are at $\cal P$ $\}$ is not
an element of reality. The mutual disturbance of the measurements,
which exists in a pre-selected and post-selected situation (even for
measurements of commuting variables), explains the cause of the
failure of the ``and" rule.

But is there a failure of the ``and" rule for weak elements of
reality?
Weak elements of reality are defined as weak values, the
outcomes of weak measurements, with the basic property  that
they
do not disturb  quantum states significantly.  Consider our first
example: two spin-1/2 particles in the EPR-Bohm state, postselected in
a state $|{\uparrow_1}_x {\uparrow_2}_y \rangle $. We have found that
there are elements of reality (which are also weak elements of
reality) $\{{\sigma_1}_y = -1\}$ and $\{{\sigma_2}_x = -1\}$.
Clearly, weak
simultaneous measurements of ${\sigma_1}_y$ and ${\sigma_2}_x$ will
not disturb  each other (while strong measurements
certainly will).  However, this does not mean that the ``and" rule
holds for weak elements of reality.  It only means that weak elements
of reality can be measured (on the pre- and post-selected ensemble)
simultaneously.  Weak measurement of ${\sigma_1}_y$ together with
weak measurement of ${\sigma_2}_x$ {\it are not} equivalent to  weak
measurement of ${\sigma_1}_y$ and ${\sigma_2}_x$.  The latter, in
fact, is not well defined.  We have to specify the two-particle
operator to be measured, and then to go to the weak limit.  For
example, the product is one such two-particle operator and since
we proved that the product rule fails in this case, the ``and" rule
must fail too. (One can see an analogy with the necessity of
specifying the operator measured in boxes 2 and 3 for
defining
the number of particles in the box 1 of our three-box
example\rlap.$^{15}$)

Even more clearly, we can see the failure of the ``and" rule for weak
elements of reality by reconsidering Hardy's example.  There
are
elements of reality $\{$electron at $\cal P\}$ and
$\{$positron at $\cal P\}$.
Weak independent measurements of the number of electrons at $\cal P$
and the number of positrons at $\cal P$ will yield the number 1 for
both. But
weak measurement of $\{$electron and positron at $\cal P\}$,
i.e. weak measurement of the number of created photons, will yield 0.

Although the failure of the ``and" rule may suggest
 a version of quantum logic, we  do not propose such a resolution.
We prefer to keep the standard logic of propositions with no
failure of the ``and" rule for propositions:
if A is true and B is
true, then
$\{$A and B$\}$ is also true. Nature, however, is described by
quantum elements of reality  which do not obey our classical
intuition.
Introduction of such elements of reality helps us construct
 a Lorentz-invariant description of the evolution
 of quantum systems\rlap.$^{21-22}$

\vskip .4cm
\noindent
{\bf 10. Acknowledgements}
\vskip .1cm

It is a pleasure to thank Sandu Popescu and Daniel Rohrlich for very
helpful discussions.  The research was supported in part by grant
425/92-1 of the the Basic Research Foundation (administered by the
Israel Academy of Sciences and Humanities).

\vskip .4cm
\noindent
\centerline{REFERENCES}
\vskip .15cm

\item {1.} A. Einstein, B. Podolsky, and N. Rosen, Phys.  Rev. {\bf
47}, 777 (1935).
\item {2.} J.S. Bell, Phys. {\bf 1}, 195 (1964).
\item {3.} I. Pitowsky, Phys. Lett. A {\bf 156}, 137 (1991).
\item {4.} R. Clifton, C. Pagonis, and I. Pitowsky, Philosophy of
Science Association Vol.  I, p. 114 (1992).
\item {5.} L. Hardy, Phys.  Rev.  Lett. {\bf 68}, 2981 (1992).
\item {6.} R. Clifton and P. Niemann, Phys.  Lett.  A {\bf 166},
177 (1992).
\item {7.} L. Vaidman, Phys.  Rev.  Lett. {\bf 70}, 3369 (1993).
\item {8.} A. Fine and P. Teller, Found.  Phys. {\bf8}, 629 (1978).
\item {9.} A. Peres, Phys. Lett. A {\bf 151}, 107 (1990).
\item {10.} A. Peres, Found.  Phys. {\bf 22}, 357 (1992).
\item {11.} J. von Neumann,  {\it Mathematical Foundations of Quantum
Theory} (Princeton, New Jersey:  Princeton University Press, 1983).
\item {12.} Y. Aharonov, D.Z. Albert, and L. Vaidman, Phys.  Rev. D
{\bf 34},
 1805 (1986).
\item {13.}  Y. Aharonov, P.G.  Bergmann, and J.L.  Lebowitz, Phys.
Rev.  B {\bf 134}, 1410 (1964).
\item {14.} Y.Aharonov and L. Vaidman, J. Phys.  A {\bf 24}, 2315
(1991).

\item {15.} D.Z. Albert, Y. Aharonov, and S. D'Amato,  Phys.  Rev.
Lett. {\bf 54}, 5 (1985).

\item {16.}  M. Redhead, {\it Incompleteness, Nonlocality, and
Realism} (Clarendon, Oxford, 1987) p. 72.
\item {17.} D.M.  Greenberger, M. Horne, A. Zeilinger, in {\it Bell's
Theorem, Quantum Theory, and Conception of the Universe}, edited by M.
Kafatos (Kluwer, Dordrecht, 1989) p. 84.
\item {18.} A. Elitzur and L. Vaidman, Found.  Phys.  (to be
published).
\item {19.} N.D.  Mermin, Phys.  Rev.  Lett. {\bf 65}, 3373 (1990)).
\item {20.} Y. Aharonov and L. Vaidman, Phys.  Rev.  A {\bf 41}, 11
(1990).
\item {21.} L. Vaidman, {\it PhD Thesis}, Tel-Aviv University 1987.
\item {22.} Y. Aharonov and D. Rohrlich, in {\it Quantum Coherence},
edited by J.S. Anandan (World-Scientific, 1990) p. 221.

 \bye